\documentclass[preprintnumbers,amsmath,amssymb,floatfix,12pt,prd,superscriptaddress,nofootinbib]{revtex4}
\usepackage{graphicx}
\usepackage{epsfig}
\usepackage{bm}
\usepackage{amsfonts}
\usepackage{subfigure}

\def\bea{\begin{eqnarray}}
\def\eea{\end{eqnarray}}

\begin{document}
\begin{center}
\LARGE { \bf  Warm-Intermediate Inflationary Universe Model with Viscous Pressure in High Dissipative Regime
  }
\end{center}
\begin{center}
{\bf M. R. Setare\footnote{rezakord@ipm.ir} \\  V. Kamali\footnote{vkamali1362@gmail.com}}\\
 { Department of Science, University of Kurdistan,
Sanandaj, IRAN.}
 \\
 \end{center}
\vskip 3cm

\begin{center}
{\bf{Abstract}}\\
Warm inflation model with bulk viscous pressure  in the context of "intermediate inflation" where the cosmological scale factor  expands as $a(t)=a_0\exp(At^f)$, is studied. The characteristics  of this model in slow-roll approximation and in high dissipative regime  are presented in two cases:  1- Dissipative parameter $\Gamma$ as a function of scalar field $\phi$ and bulk viscous coefficient $\zeta$ as a function of energy density $\rho$.
2- $\Gamma$ and $\zeta$ are constant parameters. Scalar, tensor perturbations and spectral indices for this scenario are obtained.
The cosmological parameters appearing in the present  model are constrained by recent observational
data (WMAP7).
 \end{center}

\newpage

\section{Introduction}
Big Bang model has many long-standing problems (horizon,
flatness,...). These problems are solved in a framework of
inflationary universe models \cite{1-i}. Scalar field as a source
of inflation provides a causal interpretation of the origin of
the distribution of large scale structure, and also observed anisotropy
of cosmological microwave background (CMB) \cite{6}. Standard
models for inflationary universe are divided
into two regimes, slow-roll and reheating epochs. In slow-roll
period, kinetic energy remains small compared to potential
term. In this period, all interactions between scalar fields
(inflatons) and  other fields are neglected and as a result the universe
inflates. Subsequently, in reheating poch, the kinetic energy
is comparable to the potential energy that causes inflaton to begin an
oscillation around  minimum of the potential while losing their
energy to other fields present in the theory. After this period, the universe is filled with radiation.\\
In warm inflationary models radiation production occurs during inflationary period and
reheating is avoided \cite{2}. Thermal fluctuations may be
obtained during warm inflation. These fluctuations could play a
dominant role to produce initial fluctuations which are necessary
for Large-Scale Structure (LSS) formation. Density
fluctuation arises from thermal rather than quantum fluctuation
\cite{3-i}. Warm inflationary period ends when the universe stops
inflating. After this period the universe enters in radiation
phase smoothly \cite{2}. Finally, remaining inflatons or dominant
radiation fields created the matter components of the universe.
In warm inflation models, for simplicity, the particles which are created by the inflaton decay are considered as massless particles (or radiation). The existence of massive particles in the inflationary fluid model as a new model of inflation has been considered in Ref.\cite{4-i}, perturbation parameters of this model have been presented in Ref.\cite{5}. In this scenario the existence of massive particles may be altered the dynamics of the inflationary universe  by modification the fluid pressure. Decay of the massive particles within the fluid is an entropy-producing scalar phenomenon, in other hand  "bulk viscous pressure" has entropy-producing property. Therefore the decay of particles  may be considered by a bulk viscous pressure $\Pi=-3\zeta H$ \cite{1} where $H$ is Hubble parameter and $\zeta$ is phenomenological coefficient of bulk viscosity. This coefficient is positive-definite by the second law of thermodynamics and depends on the energy density of the fluid.\\ In this work we would like to consider the
 warm inflationary universe with bulk viscous pressure  in the particular scenario
"intermediate inflation" which is denoted by scale factor
$a(t)=a_0\exp(At^f), 0<f<1$ \cite{5-i}. The expansion of this
model is faster than power-law inflation ($a=t^p; p>1$), but
slower than standard de sitter inflation ($a=\exp(Ht)$).\\
In term of string/M-theory \cite{v1}, if the high order curvature corrections to
Einstein-Hilbert action are proportional to the Gauss-Bonnet(GB)
term, we obtain a free-ghost action. The GB term is the leading
order of the $\alpha$ (inverse string tension) expansion to the
low-energy string effective action \cite{v2}. This kind of theory
is applied for study of initial singularity problem \cite{v3},
black hole solutions \cite{v4} and the late time universe
acceleration \cite{v5}. The coupling of GB with dynamical
dilatonic scalar field in $4D$ dark energy model leads to an
intermediate form for scale factor, where $f=\frac{1}{2}$,
$A=\frac{2}{8\pi G n}$
with a constant parameter "n" \cite{v1}. Intermediate inflation model may be derived from an effective theory at low dimensions of a fundamental string theory. Therefore, the study of intermediate
inflationary model is motivated by string/M-theory \cite{v1}. In the other hand, It has been shown that, there are  eight possible asymptotic solutions for cosmological dynamics \cite{v11}. Three of these solutions have non-inflationary scale factor and another three one's of solutions give de Sitter (with scale factor $a(t)=a_0\exp(H_0 t)$), power-low (with scale factor $a(t)=t^p, p>1$), inflationary expansions. Two cases of these solutions have asymptotic expansion
 with scale factor($a=a_0\exp(A(\ln t)^{\lambda})$, which is named "logamediate inflation" and finally  intermediate inflation ($a(t)=a_0\exp(At^f), 0<f<1$).
The warm inflation with bulk viscous  has been studied
in Refs. \cite{4-i} and \cite{5}. Intermediate scale factor has been used for non-viscous warm inflation models \cite{new1}. To the best of our knowledge, the warm inflation model with viscous pressure
in the context of intermediate inflation has not been yet
studied. In this paper we will study warm inflationary universe model with bulk viscous pressure  in the context of intermediate inflation. The paper is organized as: In section II,
we give a brief review about  warm inflationary universe model with bulk viscous pressure in high dissipative regime.
In section III, we consider high dissipative warm-intermediate
inflationary phase in two cases: 1- Dissipative parameter $\Gamma$ as a
function of  field $\phi$ and viscous coefficient $\zeta$ as a function of energy density of the inflation fluid. 2- Constant dissipative
parameters $\Gamma$ and constant viscous coefficient $\zeta$ .  In this section we also,
investigate the cosmological perturbations for our model. Finally in section IV,  we present
a conclusion.
\section{The model}
Warm inflation model in a spatially flat Friedmann Robertson Walker (FRW) universe which is filled with a scalar field $\phi$ and an imperfect fluid is studied. Scalar field $\phi$ or inflaton has energy density $\rho_{\phi}=\frac{1}{2}\dot{\phi}^2+V(\phi)$. Imperfect fluid is a mixture of matter and radiation with adiabatic index $\gamma$, energy density $\rho=Ts(\phi,T)$ ($T$ is the temperature  and $s$ is the entropy density of the imperfect fluid \cite{mm-1}.) and total pressure $P+\Pi$. $\Pi=-3\zeta H $ is viscous pressure\cite{1},  where $\zeta$ is phenomenological coefficient of bulk viscosity. Friedmann equation of this model is
\begin{eqnarray}\label{1}
3H^2=\frac{\dot{\phi}^2}{2}+V(\phi)+\rho
\end{eqnarray}
where we choose $c=\hbar=8\pi G=1$.
Inflation field $\phi$ decays into the imperfect fluid with rate $\Gamma$, so the conservation equation of fluid and inflaton field have these forms
\begin{eqnarray}\label{2}
\dot{\rho}+3H(\rho+P+\Pi)=\dot{\rho}+3H(\gamma\rho+\Pi)=\Gamma\dot{\phi}^2
 \end{eqnarray}
and
\begin{eqnarray}\label{3}
\dot{\rho}_{\phi}+3H(\rho_{\phi}+P_{\phi})=-\Gamma \dot{\phi}^2\Rightarrow \ddot{\phi}+(3H+\Gamma)\dot{\phi}=-V'
\end{eqnarray}
respectively. Where $V'=\frac{dV}{d\phi}$, $P=(\gamma-1)\rho$. Dissipation term denotes the inflaton decay into the imperfect fluid in the inflationary epoch.
We would like to express the evolution equation (\ref{2})  in terms of entropy
density $s(\phi,T)$. This parameter is defined by a thermodynamical relation \cite{mm-1}
\begin{equation}\label{}
s(\phi,T)=-\frac{\partial f}{\partial T}=-\frac{\partial V}{\partial T}.
\end{equation}
$f$ is Helmholtz free energy which is defined by
\begin{equation}\label{}
f=\rho_{T}-Ts=\frac{1}{2}\dot{\phi}^2+V(\phi)+\rho-Ts
\end{equation}
Free energy $f$  is dominated by  the thermodynamical potential $V(\phi, T)$ in slow-roll limit. The total energy density and total pressure are given by
\begin{eqnarray}\label{}
\rho_T=\frac{1}{2}\dot{\phi}^2+V(\phi)+Ts~~~~~~~~~~~~~~~~\\
\nonumber
P_T=\frac{1}{2}\dot{\phi}^2-V(\phi)+(\gamma-1)Ts+\Pi
\end{eqnarray}
The viscous pressure for an expanding universe is negative ($\Pi=-3\zeta H$), therefore this term acts to decrease the total pressure. Using Eq.(\ref{2}), we can find the entropy density evolution for our model as
\begin{eqnarray}\label{nn-1}
T\dot{s}+3H(\gamma Ts+\Pi)=\Gamma\dot{\phi}^2
\end{eqnarray}
In the above equation, it is assumed that $\dot{T}$ is negligible.
For a quasi-equilibrium high temperature thermal bath as an inflationary fluid, we have $\gamma=\frac{4}{3}$. The bulk viscosity effects may be read from above equation. Thus bulk viscous pressure $\Pi$ as a negative quantity, enhances the source of entropy density on the RHS of the  evolution equation (\ref{nn-1}). Therefore,  energy density of radiation  and entropy density increase by the bulk viscosity pressure $\Pi$ (see FIG.1 and FIG.2).

During the inflationary phase the energy density of inflation field $\phi$ is the order of the potential, i.e. $\rho_{\phi}\sim V(\phi),$ and this energy density dominates over the energy of imperfect fluid, i.e. $\rho_{\phi}>\rho,$ this limit is called stable regime \cite{mm-1}. So the Friedmann equation (\ref{1}) reduces to
\begin{eqnarray}\label{4}
3H^2=V(\phi)
\end{eqnarray}
In slow-roll limit, it is assumed that $\dot{\phi}^2\ll V(\phi),$ and $\ddot{\phi}\ll(3H+\Gamma)\dot{\phi}$ \cite{3}. When the decay of the inflaton to imperfect fluid is quasi-stable, we have $\dot{\rho}\ll 3H(\gamma\rho+\Pi),$ and $\dot{\rho}\ll\Gamma\dot{\phi}^2$. Therefore the equations (\ref{2}) and (\ref{3}) are reduced to
\begin{eqnarray}\label{5}
3H(1+r)\dot{\phi}=-V'
\end{eqnarray}
and
\begin{eqnarray}\label{6}
\rho\simeq\frac{r\dot{\phi}^2-\Pi}{\gamma}
\end{eqnarray}
where $r=\frac{\Gamma}{3H}$. In the present work we will restrict our analysis in high dissipative regime, i.e. $r\gg 1,$ where the dissipation coefficient $\Gamma$ is much greater than $3H$. The reason of this choice is as following. In weak dissipative, i.e. $r\ll 1$, the expansion of the universe in the inflationary era disperses the decay of the inflaton. There is a little  chance for interaction between the sectors of the inflationary fluid,
therefore  we do not have non-negligible bulk viscosity. Warm inflation in high and weak dissipative regimes for a model without bulk viscous pressure have been studied in Refs. \cite{2} and \cite{3}  respectively. Dissipation parameter $\Gamma$ may be constant or a positive function of inflaton $\phi$ and temperature $T$ by the second law of thermodynamics. There are some specific forms for the dissipative coefficient, with the most common which are found in the literatures being the $\Gamma\sim T^3$ form \cite{mm-1},\cite{2nn},\cite{3nn},\cite{4nn}.  In some works $\Gamma$ and potential of the inflaton have the same form \cite{4}. In Ref.\cite{5}, perturbation parameters for warm inflationary model with viscous pressure have  obtained where $\Gamma=\Gamma(\phi)=V(\phi)$ and $\Gamma=\Gamma_0=const$.
In this work we will study the intermediate warm inflation with viscous pressure in high dissipative regime for these  two cases. \\
Slow-roll parameters $\epsilon$ and $\eta$ in high dissipative regime are given by \cite{5}
\begin{eqnarray}\label{7}
\epsilon\equiv-\frac{\dot{H}}{H^2}=\frac{1}{2r}[\frac{V'}{V}]^2
  \end{eqnarray}
and
\begin{eqnarray}\label{8}
\eta\equiv-\frac{\ddot{H}}{H\dot{H}}=\frac{1}{r}[\frac{V''}{V}-\frac{1}{2}(\frac{V'}{V})^2]
\end{eqnarray}
respectively.
We consider potentials of the form \cite{po}
\begin{eqnarray}\label{po}
V=V_0\phi^n
\end{eqnarray}
where $V_0$ is a constant. We restrict the model in the region $\phi>0$ where the above potential is positive for all $n$. The  slow-roll condition ($\eta,\epsilon\ll 1$), are satisfied when $\phi^2$ is much greater than $\frac{n^2}{r}$. Therefore these potentials are classified as "large field" models \cite{la}.

By using Eqs.(\ref{5}), (\ref{6}) and (\ref{7}) in slow-roll limit, a relation between the energy densities $\rho_{\phi}$ and $\rho$ is obtained as
\begin{eqnarray}\label{9}
\rho=\frac{1}{\gamma}[\frac{2}{3}\epsilon\rho_{\phi}-\Pi]
\end{eqnarray}
Using inflation condition, i.e. $\ddot{a}>1,$ or equivalently $\epsilon<1,$ and above equation, warm inflation epoch with  viscose pressure  could take place when
\begin{eqnarray}\label{10}
\rho_{\phi}>\frac{3}{2}[\gamma\rho+\Pi]
\end{eqnarray}
Our warm inflation model comes to close when $\rho_{\phi}\simeq\frac{3}{2}[\gamma\rho+\Pi]$.
The number of e-folds in high dissipative regime is given by
\begin{eqnarray}\label{11}
N(\phi)=-\int_{\phi_i}^{\phi_f} r\frac{V}{V'}d\phi
\end{eqnarray}
where $\phi_i$ and $\phi_f$ are inflaton at the begining and end of inflation, respectively.
\section{Intermediate  inflation}
In this section we will study high dissipative warm inflation with viscous pressure in the context of intermediate inflation. The scale factor of intermediate inflation follows the law
\begin{eqnarray}\label{12}
a(t)=a_0\exp(At^f),~~~~~0<f<1
\end{eqnarray}
where $A$ is a positive constant with unit $m_{p}^f$. We consider our model in two cases \cite{5}: 1- $\Gamma$ is a function of scalar field $\phi$  and $\zeta$ is a function of energy density $\rho$. 2- $\Gamma$ and $\zeta$ are constant parameters.
\subsection{$\Gamma=\Gamma(\phi)=V(\phi)$, $\zeta=\zeta(\rho)=\zeta_1\rho$ case}
By using Eqs.(\ref{4}) and (\ref{5}) in this case, we get the scalar field $\phi$, Hubble parameter $H(\phi)$ and potential $V(\phi)$ as
\begin{eqnarray}\label{13}
\phi=2\sqrt{2(1-f)t}
\end{eqnarray}

\begin{eqnarray}\label{14}
H(\phi)=fA(\frac{\phi}{2\sqrt{2(1-f)}})^{2f-2}
\end{eqnarray}

and the potential in this case has form (\ref{po}), where 
\begin{eqnarray}\label{}
n=4f-4~~~~~~~~~~~~~~V_0=3(fA)^2(2\sqrt{2(1-f)})^{-n}
\end{eqnarray}

 Energy density $\rho$ is obtained from Eq.(\ref{6}) as
\begin{eqnarray}\label{aa}
\rho=\frac{2fA(1-f)[\frac{\phi}{2\sqrt{2(1-f)}}]^{2f-4}}{\gamma-3\zeta_1fA[\frac{\phi}{2\sqrt{2(1-f)}}]^{2f-2}}
\end{eqnarray}
Bulk viscous relation only holds for small deviations
from equilibrium, so we consider this limitation as
\begin{equation}\label{lim}
\mid\Pi\mid\ll\rho
\end{equation}
From two above equations the region of $\phi$ is presented by
\begin{equation}\label{lim1}
\frac{\phi}{2\sqrt{2(1-f)}}\gg(3fA\zeta_1)^{\frac{1}{2(1-f)}}
\end{equation}
The entropy density in terms of  inflaton field $\phi$ may be obtained from above equation
\begin{equation}\label{}
Ts=\frac{2fA(1-f)[\frac{\phi}{2\sqrt{2(1-f)}}]^{2f-4}}{\gamma-3\zeta_1fA[\frac{\phi}{2\sqrt{2(1-f)}}]^{2f-2}}
\end{equation}
In FIG.1,  we plot the entropy density in terms of scalar field.
\begin{figure}[h]
\centering
  \includegraphics[width=10cm]{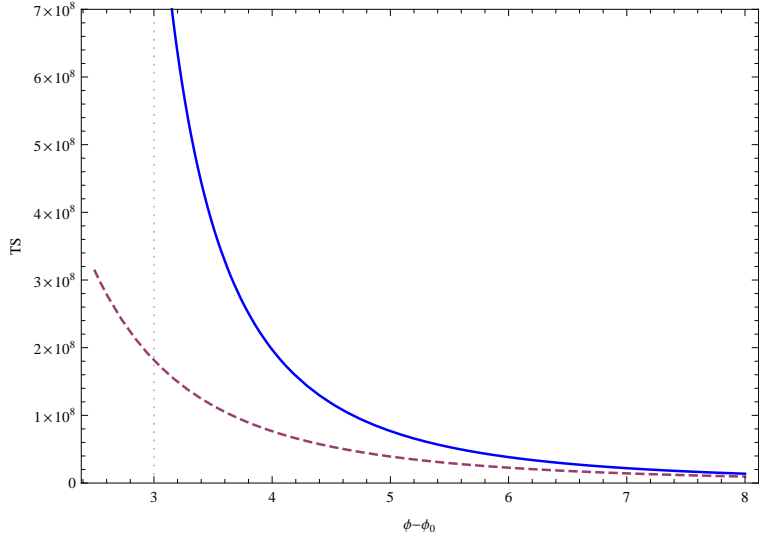}
  \caption{We plot the entropy density $s$ in terms of inflaton $\phi$ where, $\Pi=0$ by dashed curve and $\Pi=-3\zeta_1\rho H$ by blue curve ($T=5.47\times 10^{-5}$, $f=\frac{1}{2}, \gamma=1.5, A=5.01\times 10^{8}, \zeta_1=0.2\times 10^{-8}$)}
 \label{fig:F1}
\end{figure}
Using Eq.(\ref{7}) slow-roll parameter $\epsilon$ in terms of inflaton $\phi$ is given by
\begin{eqnarray}\label{16}
\epsilon=\frac{1-f}{fA}(\frac{\phi}{2\sqrt{2(1-f)}})^{-2f}
\end{eqnarray}

Likewise, using (\ref{8}) the slow-roll parameter $\eta$ has this form
\begin{eqnarray}\label{17}
\eta=\frac{3-2f}{2fA}(\frac{\phi}{2\sqrt{2(1-f)}})^{-2f}
\end{eqnarray}
The inflation condition $\ddot{a}>0$ (or equivalently $\epsilon<1$) for this example is satisfied when
\begin{eqnarray}\label{}
\phi^2>8(1-f)(\frac{1-f}{fA})^{\frac{1}{f}}
\end{eqnarray}

From equation (\ref{11}), the number of e-folds between initial and final fields $\phi_i$ and $\phi_f$ is
\begin{eqnarray}\label{}
N(\phi)=A([\frac{\phi_f}{2\sqrt{2(1-f)}}]^{2f}-[\frac{\phi_i}{2\sqrt{2(1-f)}}]^{2f})
\end{eqnarray}
We find $\phi_i$ at the begining of inflation (when $\epsilon\simeq 1$)
\begin{eqnarray}\label{18}
\phi_i=2\sqrt{2(1-f)}(\frac{1-f}{fA})^{\frac{1}{2f}}
\end{eqnarray}

So, we could find the value of $\phi_f$ in terms of $N$, $A$ and $f$ as
\begin{eqnarray}\label{19}
\phi_f=2\sqrt{2(1-f)}[\frac{N}{A}+\frac{1-f}{fA}]^{\frac{1}{2f}}
\end{eqnarray}

Now, we study the scalar and tensor perturbation spectrums for our model in this case ($\Gamma=V,\zeta=\zeta_1\rho$). The power-spectrum of the curvature perturbation have the form \cite{5}
\begin{eqnarray}\label{20}
P_R=\frac{1}{2\pi^2}\exp(-2\Im(\phi))[\frac{T_r}{\epsilon\sqrt{rV^3}}]
\end{eqnarray}

The amplitude of tensor perturbation which could produce gravitational waves during inflation is given by
\begin{eqnarray}\label{21}
A_{g}^2=2(\frac{H}{2\pi})^2\coth[\frac{k}{2T}]\simeq\frac{f^2A^2}{2\pi^2}(\frac{\phi}{2\sqrt{2(1-f)}})^{4f-4}\coth[\frac{k}{2T}]
\end{eqnarray}

In the above equations the temperature of thermal background of gravitational wave has found in extra factor $\coth[\frac{k}{2T}]$ and
\begin{eqnarray}\label{22}
\Im(\phi)=-\int{\{\frac{\Gamma'}{3Hr}+\frac{3}{8}[1-((\gamma-1)+\frac{\Pi}{\zeta}\frac{d\zeta}{d\rho})\frac{\Gamma'V'}{9\gamma rH^2}]\frac{V'}{V}}\}d\phi\\
\nonumber
=-\frac{11}{2}(1-f)\ln(\phi)+\alpha[\frac{\phi-\phi_0}{2\sqrt{2(1-f)}}]^{2f-4}\\
\nonumber
+\beta(\frac{\phi}{2\sqrt{2(1-f)}})^{4f-6}~~~~~~~~~~~~~~~~~~~~~~~~~~~~~~~
\end{eqnarray}

where $\alpha=\frac{12(\gamma-1)(1-f)^2fA}{(2-f)\gamma}$ and $\beta=-\frac{18\zeta_1(1-f)^2}{\gamma(3-2f)}$.
From Eqs.(\ref{20}) and (\ref{21}) in high dissipative regime the tensor-to-scalar ratio is given by
\begin{eqnarray}\label{23}
R(k_0)=(\frac{A_g^2}{P_R})|_{k=k_0}=\frac{2}{3}[\frac{\epsilon\sqrt{rV^5}}{T_r}]\exp(2\Im(\phi))\coth[\frac{k}{2T}]|_{k=k_0}
\end{eqnarray}

Using seven-year Wilkinson microwave anisotropy probe (WMAP7) observational data we find an upper bound for $R=0.21<0.36$ \cite{6}. Spectral indices $n_g$ and $n_s$ in the present case are
\begin{eqnarray}\label{24}
n_g=\frac{d}{d\ln k}\ln[\frac{A_g^2}{\coth[\frac{k}{2T}]}]=-2\epsilon=\frac{2(f-1)}{fA} (\frac{\phi}{2\sqrt{2(1-f)}})^{-2f}
\end{eqnarray}

and
\begin{eqnarray}\label{25}
n_s=1-\frac{4}{25}\frac{d\ln P_R}{d\ln k}\approx 1-[\epsilon+2\eta+\sqrt{\frac{2\epsilon}{r}}[2\Im'(\phi)-\frac{r'}{2r}]]\\
\nonumber
\simeq 1-(2\eta-5\epsilon)=1-\frac{3f-2}{fA}(\frac{\phi}{2\sqrt{2(1-f)}})^{-2f}
\end{eqnarray}
where $d\ln k(\phi)=-dN(\phi)$. Since $0<f<1$ we obviously see that the
Harrison-Zeldovich spectrum  (i.e. $n_s=1$) occurs for $f=\frac{2}{3}$
which  agrees with  and non-viscous inflation models
\cite{5-i}, \cite{17}, \cite{18}. $n_s>1$ is equivalent to $f<\frac{2}{3}$  and
$n_s<1$ is equivalent to $f>\frac{2}{3}$. Using the limitation of $\phi$ (\ref{lim1}) in this case, which is given by the condition (\ref{lim}), and using Eq.(\ref{25}), we could find nearly scale invariant spectrum ($n\simeq 1$) for all $f,$ which agrees with WMAP7 observational data \cite{6}.
Running of the scalar spectral index is an important cosmological parameter which may be obtained by WMAP7 data
\begin{eqnarray}\label{26}
\alpha_s=\frac{dn_s}{d\ln k}=-\frac{dn_s}{d\phi}\frac{d\phi}{dN (\phi)}=-\sqrt{\frac{2\epsilon}{r}}[\epsilon'+2\eta']\\
\nonumber
-\frac{\epsilon}{r}[(\frac{\epsilon'}{\epsilon}-\frac{r'}{r})[2\Im'-\frac{r'}{2r}]+[4\Im''-(\ln r)'']]\\
\nonumber
\simeq \frac{2-3f}{2\sqrt{2(1-f)}}\frac{V'}{rV}(\frac{\phi}{2\sqrt{2(1-f)}})^{-2f-1}~~~~~~
\end{eqnarray}
In term of WMAP7 results $\alpha_s$ is approximately $-0.038$ \cite{6}. In the next subsection we will consider the specific case in which the dissipative parameter $\Gamma$ and coefficient of bulk viscosity $\zeta$ are constant parameters.
\subsection{$\Gamma=\Gamma_0$, $\zeta=\zeta_0$ case }
Where $\Gamma=\Gamma_0$ and $\zeta=\zeta_0$ and by using Eqs.(\ref{4}) and (\ref{5}) we get
 \begin{eqnarray}\label{27}
 \phi=\varpi t^{f-\frac{1}{2}}
\end{eqnarray}
where $\varpi=\frac{2fA}{2f-1}\sqrt{\frac{6(1-f)}{\Gamma_0 }}$.
The effective potential and Hubble parameters in this case are obtained as
\begin{eqnarray}\label{28}
H=fA(\frac{\phi}{\varpi})^{\frac{2f-2}{2f-1}}
\end{eqnarray}
and the potential in this case has form (\ref{po}), where
\begin{eqnarray}\label{}
n=\frac{4f-4}{2f-1}~~~~~~~~~~~~~~V_0=3f^2A^2(\varpi)^{-n}
\end{eqnarray}

Viscous pressure $\Pi$ and energy density $\rho$  are obtained from Eqs.(\ref{5}), (\ref{6}) and (\ref{28})
 \begin{eqnarray}\label{}
\Pi=-3\zeta H=-3fA\zeta_0(\frac{\phi}{\varpi})^{\frac{2f-2}{2f-1}}~~~~~~~~~\\
\nonumber
\rho=\frac{fA}{\gamma}[2(1-f)(\frac{\phi}{\varpi})^{\frac{2f-4}{2f-1}}+3\zeta_0(\frac{\phi}{\varpi})^{\frac{2f-2}{2f-1}}]
\end{eqnarray}
Using the above equation and Eq.(\ref{lim}) we could constrain the scalar field  $\phi$,  as:
\begin{eqnarray}\label{lim2}
f<\frac{1}{2}~~\Rightarrow~~\phi\gg\varpi(\frac{\zeta_0(\gamma-1)}{2(1-f)})^{\frac{1-2f}{2}}\\
\nonumber
f>\frac{1}{2}~~\Rightarrow~~\phi\ll\varpi(\frac{\zeta_0(\gamma-1)}{2(1-f)})^{\frac{1-2f}{2}}
\end{eqnarray}
We can find the entropy density $s$ in terms of scalar field
\begin{equation}\label{}
Ts=\frac{fA}{\gamma}[2(1-f)(\frac{\phi}{\varpi})^{\frac{2f-4}{2f-1}}+3\zeta_0(\frac{\phi}{\varpi})^{\frac{2f-2}{2f-1}}]
\end{equation}
The entropy density and energy density in term of our model in this case increase by the bulk viscosity effect (see FIG.2).
\begin{figure}[h]
\centering
  \includegraphics[width=10cm]{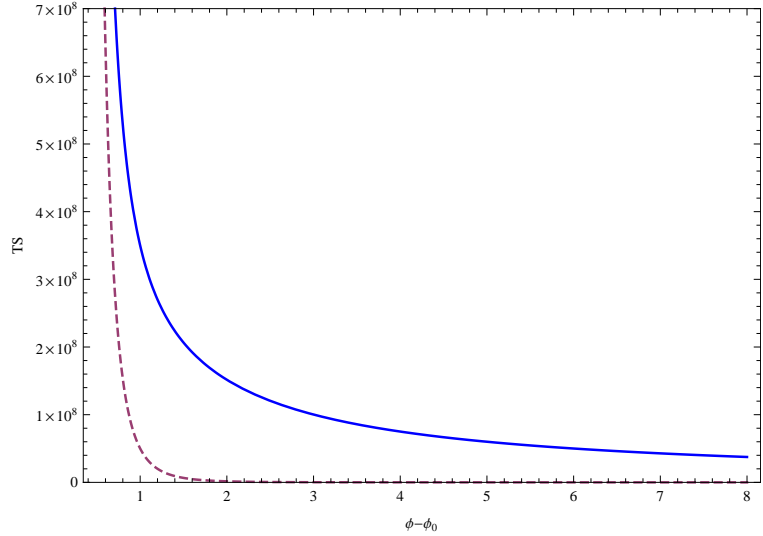}
  \caption{We plot the entropy density $s$ in terms of inflaton $\phi$ where, $\Pi=0$ by dashed curve and $\Pi=-3\zeta_0 H$ by blue curve ($T=5.47\times 10^{-5}$, $f=\frac{3}{4}, \gamma=1.5, A=2\times 10^{8}, \zeta_0=1, \Gamma_0=5.4\times 10^{17}$)}
 \label{fig:F1}
\end{figure}
Using Eqs.(\ref{7}) and (\ref{8}) the slow-roll parameters $\epsilon$ and $\eta$ in term of scalar field $\phi$  are
\begin{eqnarray}\label{30}
\epsilon\simeq\frac{1-f}{fA}(\frac{\phi}{\varpi})^{\frac{-2f}{2f-1}}\\
\nonumber
\eta\simeq\frac{1}{2fA}(\frac{\phi}{\varpi})^{\frac{-2f}{2f-1}}
\end{eqnarray}
respectively.
The inflation condition $\ddot{a}>0$ (or equivalently $\epsilon<1$) for this  case is satisfied when
\begin{eqnarray}\label{}
\nonumber
\phi^2>\varpi(\frac{1-f}{fA})^{\frac{2f-1}{2f}}
\end{eqnarray}
From Eq.(\ref{11}) the number of e-folds is found
\begin{eqnarray}\label{31}
N=A[(\frac{\phi_f}{\varpi})^{\frac{2f}{2f-1}}-(\frac{\phi_i}{\varpi})^{\frac{2f}{2f-1}}]
\end{eqnarray}
At the begining of the inflation period ($\epsilon\simeq 1$) $\phi_i=\varpi(\frac{1-f}{fA})^{\frac{2f-1}{2f}}$, so the scalar field $\phi_f$ at the end of inflation in term of the number of e-folds becomes
\begin{eqnarray}\label{32}
\phi_f=\varpi(\frac{N}{A}+\frac{1-f}{fA})^{\frac{2f-1}{2f}}
\end{eqnarray}

In the following we will find the perturbation parameters in terms of scalar field. From Eq.(\ref{22}) we have
\begin{eqnarray}\label{33}
 \Im(\phi)=-\frac{3}{8}\ln(V(\phi))
\end{eqnarray}

The spectrum of the curvature perturbation in slow-roll limit in this case ($\Gamma=\Gamma_0$, $\zeta=\zeta_0$), from above equation and equation (\ref{20}), has the form
\begin{eqnarray}\label{34}
P_{R}\approx B(\frac{\phi}{\varpi})^{\frac{2}{2f-1}}
\end{eqnarray}
where $B=\frac{T_r}{2\pi^2(1-f)\sqrt{\Gamma_0\sqrt{3}}}$. This parameter is found from WMAP7 results ($P_R=2.28\times 10^{-9}$)\cite{6}. Using Eq.(\ref{21}) and (\ref{33}) the amplitude of tensor perturbation becomes
\begin{eqnarray}\label{35}
A_g^{2}=\frac{f^2A^2}{2\pi^2}(\frac{\phi}{\varpi})^{\frac{4f-4}{2f-1}}\coth[\frac{k}{2T}]
\end{eqnarray}

From Eq.(\ref{24}) the spectral index $n_s$ is given  by
\begin{eqnarray}\label{36}
n_s=1-\frac{2}{fA}(\frac{\phi}{\varpi})^{\frac{-2f}{2f-1}}
\end{eqnarray}
For intermediate inflation  where  $0<f<1$, the scalar index $n_s$ for this example  becomes $n_s<1$. By using the limitation (\ref{lim2}) and above equation, we could obtain nearly scale invariant spectrum ($n\simeq 1$) for $f>\frac{1}{2}$.
Using Eqs.(\ref{31}) and (\ref{32}) we can re-express the above index in terms of number of e-folding
\begin{eqnarray}\label{}
n_s=1-\frac{2}{1+f(N-1)}
\end{eqnarray}
and from above equation we find the value of $f$ in terms of $N$ and $n_s$ as
\begin{eqnarray}\label{}
f=\frac{1+n_s}{(1+n_s)(N-1)}
\end{eqnarray}
Using WMAP7 observational data $n_s\simeq 0.96$ and $N=60$ as a standard benchmark we obtain $f\simeq 0.83$. This amount of $f$ is also in the region $f>\frac{1}{2}$.
From Eq.(\ref{25}) the spectral index $n_g$ becomes
\begin{eqnarray}\label{}
n_g=\frac{2(f-1)}{fA}(\frac{\phi}{\varpi})^{\frac{-2f}{2f-1}}
\end{eqnarray}
We could find the tensor-scalar ratio as
\begin{eqnarray}\label{aaa}
R=\frac{(1-f)\sqrt{\sqrt{3}\Gamma_0}f^2A^2}{T_r}(\frac{\phi}{\varpi})^{\frac{4f-6}{2f-1}}\coth[\frac{k}{2T}]\\
\nonumber
=\frac{(1-f)\sqrt{\Gamma_0\sqrt{3}}f^2A^2}{T_r}\coth[\frac{k}{2T}][\frac{2}{fA(1-n_s)}]^{\frac{4f-6}{2f-1}}
\end{eqnarray}
In FIG.3, the dependence of the tensor-to-scalar ratio on the spectrum index is shown. Three different values for parameter $\Gamma_0$ have been used in this figure, when the value $f=\frac{3}{4}$ is taken.
We note that for different values of $\Gamma_0$ which are bounded from below, $\Gamma_0>0.275$ our model is well supported by WMAP data.

\begin{figure}[h]
\centering
\includegraphics[width=10cm]{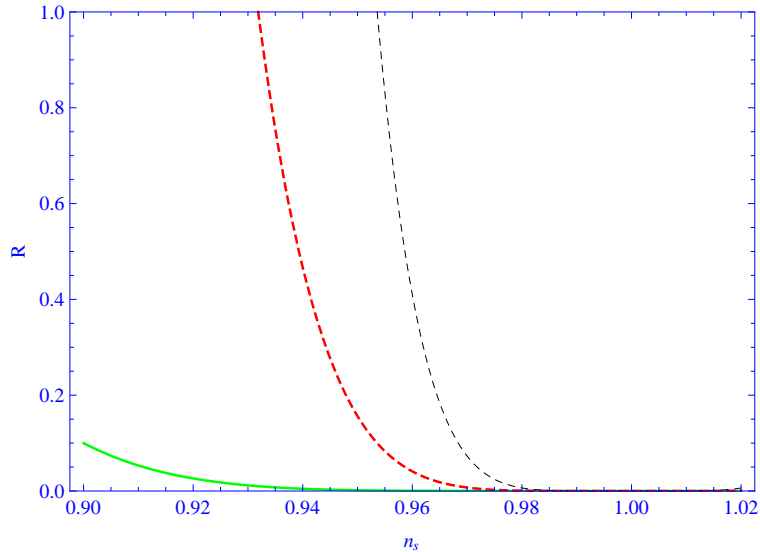}
  \caption{ We plot the evolution of the tensor-scalar ratio $r$ versus spectrum index $n_s$, for three cases: 1- $\Gamma_0=0.275$ by green line,
  2- $\Gamma=27.5$ by red dashed line, 3- $\Gamma_0=2.75\times10^4$ by black dashed line.($f=\frac{3}{4}$, $A=1,$ $T=T_r=5.47 \times 10^{-5},$ $ k=0.002 Mpc^{-1}$)
   }
 \label{fig:F3}
\end{figure}
Running of the scalar spectral index is obtained from Eq.(\ref{26})
\begin{eqnarray}\label{}
\alpha_s=\frac{dn_s}{d\ln k}=-\frac{dn_s}{d\phi}\frac{d\phi}{dN}=-\frac{4f}{r(2f-1)\varpi}\frac{V'}{rV}(\frac{\phi}{\varpi})^{\frac{-4f+1}{2f-1}}
\end{eqnarray}
This parameter may be found from WMAP7 observational data \cite{6}. Using WMAP7 data, $P_R(k_0)=\simeq 2.28\times 10^{-9}$, $R(k_0)\simeq 0.21$ and the characteristic of warm inflation $T>H$ \cite{3}, we may restrict the values of temperature to $T_r>5.47\times 10^{-5}M_4$ using Eqs.(\ref{34}), (\ref{aaa}),
(see FIG.4). We have chosen $k_0=0.002 Mpc^{-1}$ and $T\simeq T_r$. Note that, because of the bulk viscous pressure,
the radiation energy density in our model increases. Therefore the minimum value of temperature for our model ($5.47\times 10^{-5}M_4$) is bigger than
the minimum value of temperature ($3.42\times 10^{-6}M_4$ ) for the model without the viscous pressure effects \cite{6-f}.
\begin{figure}[h]
\centering
  \includegraphics[width=10cm]{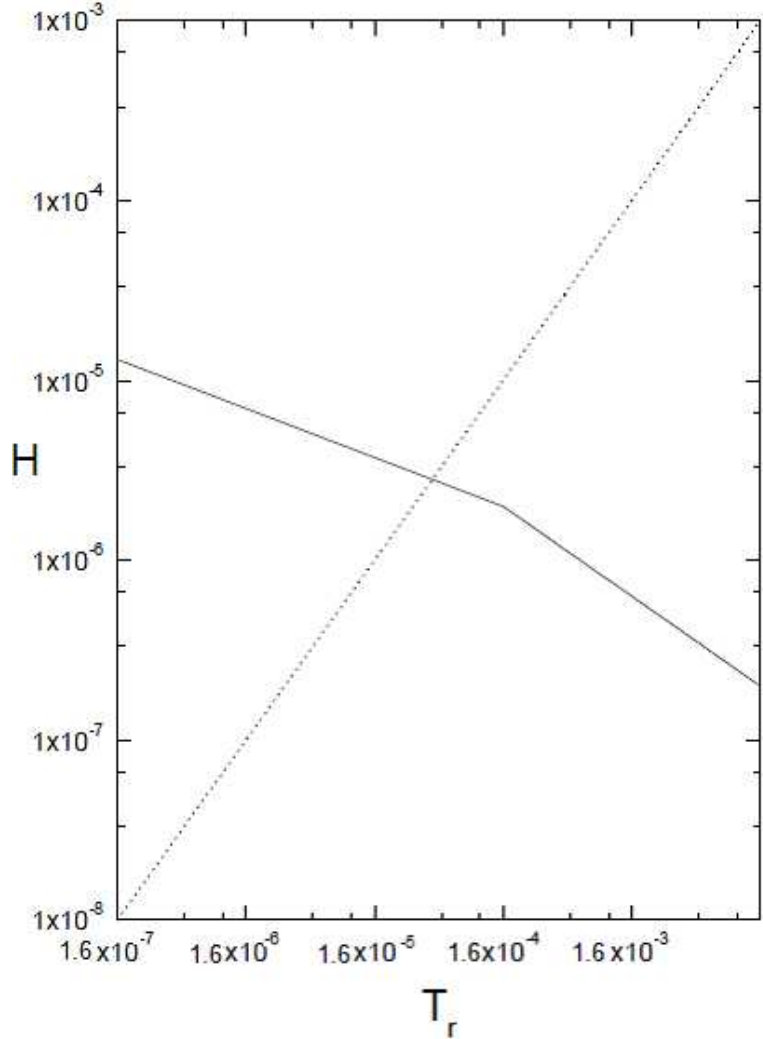}
  \caption{In this graph we plot the Hubble parameter $H$ in term of the temperature $T_r$. We can find the minimum amount of temperature $T_r=5.47\times 10^{-5}$ in order to have the necessary condition for warm inflation model ($T_r>H$). }
 \label{fig:F3}
\end{figure}
\section{Conclusion}
In this article we have investigated  the warm-intermediate
inflationary model with viscous pressure. We have studied this scenario in two
different cases of the dissipative coefficient $\Gamma$ and bulk viscous coefficient $\zeta$. Our
model have been described for $\Gamma=\Gamma_0=const$, $\zeta=\zeta_0$ and for
$\Gamma$ as a function of  field $\phi$, i.e.
$\Gamma=f(\phi)=V(\phi)$, $\zeta$ as a function of energy density $\rho$. For these two cases we have extracted the form of potential and Hubble parameters as a function of scalar field $\phi$. In $\Gamma=f(\phi)=V(\phi)$ case,
we introduced scalar field potential as $V(\phi)\propto
\phi^{4(f-1)}, 0<f<1$, but in non-viscous inflation
potential has the form $\phi^{-4\frac{1-f}{f}}$. In this case, it is
possible in the slow-roll approximation to have the
Harrison-Zeldovich spectrum of density perturbation (i.e. $n_s=1$
), provided $f$ takes the value of $\frac{2}{3}$ which
 agrees with regular inflation model with a canonical scalar
field characterized by a quasi-exponential expansion. Explicit expressions for tensor-scalar ratio $R$,  spectrum indices $n_g$ and $n_s$, running of the scalar spectral index $\alpha_s$ in slow-roll were obtained. We also have constrained these parameters by WMAP7 results.



\begin{thebibliography}{99}
\bibitem{1-i} A. Guth, "The inflationary universe: A possible solution to the horizon and flatness problems,"
 Phys. Rev. D 23, 347, (1981); A. Albrecht  and P. J. Steinhardt, "Cosmology for grand unified theories with radiatively induced
symmetry breaking," Phys. Rev. Lett. 48, 1220, (1982); A complete
description of inflationary scenarios can be found in the book by
A. Linde, "Particle physics and inflationary cosmology," (Gordon
and Breach, New York, 1990).
\bibitem{6} WMAP collaboration, E. Komatsu et al., Seven-year Wilkinson Microwave Anisotropy Probe
(WMAP) observations: cosmological interpretation, Astrophys. J. Suppl. 192 (2011) 18
[arXiv:1001.4538];
B. Gold et al., Seven-year Wilkinson Microwave Anisotropy Probe (WMAP) observations:
galactic foreground emission, Astrophys. J. Suppl. 192 (2011) 15 [arXiv:1001.4555];
D. Larson et al., Seven-year Wilkinson Microwave Anisotropy Probe (WMAP) observations:
power spectra and WMAP-derived parameters, Astrophys. J. Suppl. 192 (2011) 16
[arXiv:1001.4635].
\bibitem{2} A. Berera, Warm inflation, Phys. Rev. Lett. 75 (1995) 3218 [astro-ph/9509049];
Interpolating the stage of exponential expansion in the early universe: a possible alternative
with no reheating, Phys. Rev. D 55 (1997) 3346 [hep-ph/9612239].
\bibitem{3-i} L. M. H. Hall, I. G. Moss and A. Berera, Phys.Rev.D 69, 083525 (2004); I.G. Moss, Phys.Lett.B 154, 120 (1985); A. Berera,
Nucl.Phys B 585, 666 (2000).
\bibitem{4-i} J.P. Mimoso, A. Nunes, and D. Pav´on, Phys. Rev. D 73, 023502 (2006).
\bibitem{5}
  S. del Campo, R. Herrera and D. Pavon,
  Cosmological perturbations in warm inflationary models with viscous pressure,''
    Phys.\ Rev.\ D {\bf 75}, 083518 (2007)  [astro-ph/0703604 [ASTRO-PH]].
\bibitem{1} L. Landau and E.M. Lifshitz, Mecanique des Fluides (MIR, Moscow, 1971); K. Huang, Statistical
Mechanics (J. Wiley, New York, 1987).
\bibitem{5-i} A. Vallinotto, E. J. Copeland,
E. W. Kolb, A. R. Liddle and D. A. Steer, Phys. Rev. D 69, 103519
(2004); A. A. Starobinsky JETP Lett. 82, 169 (2005).
\bibitem{v1} A. K. Sanyal, Phys. Lett. B. 645, 1 (2007).
\bibitem{v2} T. Kolvisto and D. Mota, Phys. Lett. B 644 104 (2007); Phys. Rev. D. 75 023518 (2007).
\bibitem{v3} I. Antoniadis, J. Rizos and K. Tamvakis, Nucl. Phys. B 415 497 (1994).
\bibitem{v4} S. Mignemi and N. R. Steward, Phys. Rev. D 47 5259 (1993).
\bibitem{v5} S. Nojiri, S. D. Odintsov and M. Sasaki, Phys. Rev. D 71 123509 (2004); G. Gognola, E. Eizalde, S.
Nojiri, S. D. Odintsov and E. Winstanley, Phys. Rev. D 73 084007
(2006).
\bibitem{v11} J. D. Barrow, Class. Quantum Grav. 13, 2965 (1996).
\bibitem{new1} S.~del Campo and R.~Herrera,
  JCAP {\bf 0904}, 005 (2009)
  [arXiv:0903.4214 [astro-ph.CO]]; R.~Herrera and E.~San Martin,
  Eur.\ Phys.\ J.\ C {\bf 71}, 1701 (2011)
  [arXiv:1108.1371 [gr-qc]]; M.~R.~Setare and V.~Kamali,
  JCAP {\bf 1208}, 034 (2012)
  [arXiv:1210.0742 [hep-th]].
\bibitem{mm-1} M.~Bastero-Gil, A.~Berera, R.~Cerezo, R.~O.~Ramos and G.~S.~Vicente,
  JCAP {\bf 1211}, 042 (2012)
  [arXiv:1209.0712 [astro-ph.CO]].
\bibitem{3} A. Berera, M. Gleiser and R.O. Ramos, Strong dissipative behavior in quantum field theory,
Phys. Rev. D 58 (1998) 123508 [hep-ph/9803394]; A first principles warm inflation
model that solves the cosmological horizon/flatness problems, Phys. Rev. Lett. 83 (1999) 264
[hep-ph/9809583].
\bibitem{2nn} M.~Bastero-Gil and A.~Berera,
  Int.\ J.\ Mod.\ Phys.\ A {\bf 24}, 2207 (2009)
  [arXiv:0902.0521 [hep-ph]].
\bibitem{3nn} A.~Berera, I.~G.~Moss and R.~O.~Ramos,
  Rept.\ Prog.\ Phys.\  {\bf 72}, 026901 (2009)
  [arXiv:0808.1855 [hep-ph]].
\bibitem{4nn} M.~Bastero-Gil, A.~Berera and R.~O.~Ramos,
  JCAP {\bf 1109}, 033 (2011)
  [arXiv:1008.1929 [hep-ph]].
\bibitem{4} R. Herrera, S. del Campo and C. Campuzano, Tachyon warm inflationary universe models,
JCAP 10 (2006) 009 [astro-ph/0610339]; S. del Campo and R. Herrera, Curvaton field and intermediate inflationary universe model,
Phys. Rev. D 76 (2007) 103503 [arXiv:0710.5524].
\bibitem{po} A. Linde, Phys. Letts. B 129, 177 (1983).
\bibitem{la} B. Bassett, S. Tsujikawa and D.Wands, Rev. Mod. Phys.
78, 537 (2006).
\bibitem{17} S. del Campo, R. Herrera, and A. Toloza, "Tachyon Field in Intermediate Inflation," Phys. Rev. D79, 083507, (2009).
\bibitem{18} J. D Barrow and A. R. Liddle, Phys. Rev. D 47, R5219 (1993).
\bibitem{6-f} M. Antonella Cid, S. del Campo, R. Herrera, Warm inflation on the brane, JCAP 0710:005, (2007).
\end{thebibliography}
\end{document}